\documentclass[12pt]{article}
\makeatletter
\def\eqnarray{%
 \stepcounter{equation}%
 \let\@currentlabel=\theequation
 \global\@eqnswtrue
 \global\@eqcnt\z@
 \tabskip\@centering
 \let\\=\@eqncr
 $$\halign to \displaywidth\bgroup\@eqnsel\hskip\@centering
 $\displaystyle\tabskip\z@{##}$&\global\@eqcnt\@ne
 \hfil$\displaystyle{{}##{}}$\hfil
 &\global\@eqcnt\tw@$\displaystyle\tabskip\z@{##}$\hfil
 \tabskip\@centering&\llap{##}\tabskip\z@\cr}
\makeatother

\renewcommand{\theequation}{\thesection.\arabic{equation}}

\usepackage{amssymb}

\date{\empty}

\begin{document}

\title{$N=4$ Super-Schwarzian Theory  \\
on the Coadjoint Orbit and PSU(1,1$|$2)}

\vspace{2cm}

\author{Shogo Aoyama\thanks{Professor emeritus, e-mail: aoyama.shogo@shizuoka.ac.jp}
  \hspace{1cm}  Yuco Honda \thanks{Alumna, e-mail: yufrau@hotmail.co.jp } \\
       Department of Physics \\
              Shizuoka University \\
                Ohya 836, Shizuoka  \\
                 Japan}

\maketitle

\begin{abstract}
An $N=4$ super-Schwarzian theory is formulated by the coadjoint orbit method. It is discovered that the action has  symmetry under PSU(1,1$|$2).
\end{abstract}

\newpage

\section{Introduction}

The Schwarzian theory has been drawing a lot of attention giving a hope to be  a theoretical gateway between  
  the SYK model and the $D=2$ effective gravity. 
 The Liouville gravity, which is the simplest one for 
the $D=2$ effective gravity,  was extensively studied by  various methods  after the pioneering Polyakov's work.  (See \cite{A1, Se} for an overview of the studies at the early stage.) Among them the coadjoint orbit method proposed by Alekseev and Shatashvili  is the most geometrical\cite{Al}. 
 In a recent paper \cite{Wi} this coadjoint orbit method was revisited to study  the Schwarzian theory. 
The Hamilton structure of the Schwarzian theory was  clarified through the formulation  by the coadjoint orbit method. 

The first aim of this paper is to formulate an $N=4$ super-Schwarzian theory by means of the  coadjoint orbit method. 
  The lower symmetric cases have been  discussed in the past few years. (See \cite{Super,Mertens} 
 for instance.) But a proper account on   the  differential geometry 
  has been given only recently  in \cite{Wi}, but  for the non-supersymmetric case. 
  
  As for supersymmetrization of the coadjoint orbit method for  the Liouville gravity it is opportune to give a 
 brief summary of the studies.  
After the work \cite{Al} the coadjoint orbit method  was generalized to formulate the $(1,0)$ and $(2,0)$ supersymmetric   theories in \cite{A2} and \cite{Nieu,Del} respectively.  The left-moving sector was extended so as to admit the $(1,0)$ and $(2,0)$ superconformal symmetry. In  the right-moving sector the conformal symmetry remained  non-supersymmetric, but  the  symmetry SL(2) got promoted to OSp(2$|$1) and  OSp(2$|$2) for the respective supersymmetric  theories\cite{A3,A4}. A further extension of the coadjoint orbit method  to  the $N=(4,0)$ supersymmetric case  has not been discussed, although the $N=4$  superconformal algebra  has been known since a long time ago\cite{Eguchi}.  

 Once formulated an $N=4$ super-Schwarzian theory the second aim of this paper is  to examine if it has a symmetry further generalizing OSp(2$|$2). We show that the theory indeed has symmetry under PSU(1,1$|$2). 

The paper is organized as follows. In Section 2 we give a short summary of the coadjoint orbit method.
 In Section 3  we discuss  the $N=4$ superconformal diffeomorphism and give the $N=4$ super-Schwarzian
 derivative, which is a key element for the paper.  
Following these arguments  the coadjoint orbit method is worked out to construct 
the $N=4$ super-Schwarzian theory in Section 4. In Section 5 we show that the theory admits   
symmetry under PSU(1,1$|$2).
 Three Appendices are devoted to help the calculations in the main body of the paper.

\section{A short summary of the coadjoint orbit mothod}
\setcounter{equation}{0}

We shall start with a brief review about the general construction of the Kirillov-Kostant 2-form
 on the coadjoint orbit of a Lie-group G\cite{Al}. Let   $\frak g$ to be a Lie-algebra of  G and $\frak g^*$ the dual space of $\frak g$. An element $g\in G$ acts on  an element $a\in \frak g$ by Ad$(g)a=gag^{-1}$, while the group action on an element $X$ of the dual space  $\frak g^*$ is defined by means of an invariant quadratic form
\begin{eqnarray}
<{\rm Ad}^*(g)X,a>=<X,{\rm Ad}(g^{-1})a>. \label{1}
\end{eqnarray}
Elements $a\in \frak g$ and $X\in {\frak g}^*$ are called adjoint and coadjoint vectors respectively. We define the Kirillov-Kostant 2-form as
\begin{eqnarray}
\Omega_X={1\over 2}<X,[Y,Y]>,   \label{KK}
\end{eqnarray}
where $Y$ is a  $\frak g$-valued 1-form related with $X$ by
\begin{eqnarray}
dX={\rm ad}^*(Y)X. \label{orbit}
\end{eqnarray}
Here ${\rm ad}^*(Y)$ is the infinitesimal coadjoint action on $X$ determined according to (\ref{1}), i.e.,
\begin{eqnarray}
<{\rm ad}^*(Y)X,a>=-<X,{\rm ad}(Y)a>=-<X,[Y,a]>.  \label{invariant}
\end{eqnarray}
(\ref{orbit}) defines an orbit in the dual space ${\frak g}^*$, called the coadjoint orbit O$_X$.
 Owing to the Jacobi identity we can show  that  $dY={1\over 2}[Y,Y]$ and 
the 2-form $\Omega_X$ is closed. This 2-from is a central tool for the coadoint orbit method.

\section{The $N=4$ superconformal diffeomorphism}
\setcounter{equation}{0}

The $N=4$ superconformal group is a group of which elements are superdiffeomorphism in the 
 $N=4$ superspace. A general account of the $N$-extended superconformal group was given in \cite{Co,Scou}. 
The case of $N=4$  was studied in \cite{Ue}. Here we elaborate their arguments.   The $N=4$ superspace is described by the supercoordinates
$$
  (x, \theta_1,\theta_2,\theta^1,\theta^2) \equiv (x, \theta). \nonumber
$$
Here $x$ is a real coordinate. $\theta_a,a=1,2$, are fermionic ones, while $\theta^a,a=1,2$, their complex conjugates.  
The supercovariant derivatives are defined by
\begin{eqnarray}
D_{\theta a}&=&{\partial\over \partial\theta^a}+\theta_a{\partial\over \partial x}\equiv \partial_{\theta a}+\theta_a\partial_x, \nonumber\\
D^{\ a}_\theta&=&{\partial\over \partial\theta_a}+\theta^a{\partial\over \partial x}\equiv \partial_{\theta}^{\ a}+\theta^a\partial_x,   \label{covderiv}
\end{eqnarray}
so as to satisfy 
\begin{eqnarray}
\{D_{\theta a},D^{\ b}_\theta\}=2\delta^b_a\partial_x, \quad\quad
\{D_{\theta a}, D_{\theta b}\}=0, \quad\quad   \{D^{\ a}_\theta, D^{\ b}_\theta\}=0.   \label{algebra} 
\end{eqnarray}
We consider  $N=4$ superdiffeomorphisms 
\begin{eqnarray}
x &\longrightarrow& f(x, \theta), \quad 
\theta_a \longrightarrow \varphi_a(x, \theta),  \quad
\theta^a \longrightarrow \varphi^a(x, \theta). 
  \label{diffeo}
\end{eqnarray}
The supercovariant derivatives change as
\begin{eqnarray}
D_{\theta a}&=&D_{\theta a} f{\partial \over \partial f}
+D_{\theta a} \varphi_b{\partial\over \partial \varphi_b} +D_{\theta a}\varphi^b{\partial\over \partial \varphi^b},   \label{D1}\\
D^{\ a}_\theta &=&D^{\ a}_\theta f{\partial \over \partial f}
+D^{\ a}_\theta \varphi_b{\partial\over \partial \varphi_b} +D^{\ a}_\theta\varphi^b{\partial\over \partial\varphi^b},   \label{D2}
\end{eqnarray}
by the chain rule. 
Impose the chirality conditions 
\begin{eqnarray}
D_{\theta a}\varphi_b=0, \quad\quad D^{\ a}_\theta\varphi^b=0, 
 \label{cond1}
\end{eqnarray}
and the superconformal conditions 
\begin{eqnarray}
D_{\theta a} f=\varphi_b D_{\theta a} \varphi^b,\quad\quad D^{\ a}_\theta f= \varphi^b D^{\ a}_\theta\varphi_b.  
 \label{cond2}
\end{eqnarray}
Then (\ref{D1}) and (\ref{D2}) become  supercovariant derivatives   as
\begin{eqnarray}
D_{\theta a}&=& D_{\theta a}\varphi^b D_{\varphi b}\equiv D_{\theta a}\varphi^b({\partial \over \partial \varphi^b} +\varphi_b{\partial\over \partial f}), \nonumber\\
D^{\ a}_\theta&=& D^{\ a}_\theta{\varphi_b} D_{\varphi}^{\ b}\equiv D^{\ a}_\theta\varphi_b({\partial \over \partial \varphi_b} +\varphi^b{\partial\over \partial f}).
\nonumber 
\end{eqnarray}
When the supercovariant derivatives satisfy these transformation properties, the transformations  in (\ref{diffeo}) are called superconformal diffeomorphisms.  
Elements of the superconformal group consist of such  diffeomorphisms.

A superfield with weight 0, denoted by  $\Psi_0(x,\theta)$, transforms as 
\begin{eqnarray}
\Psi_0(x,\theta)  
 \longrightarrow \Psi_0(f(x,\theta)),\varphi(x,\theta)), \nonumber
\end{eqnarray}
by the superconformal diffeomorphisms. 
Infinitesimally it reads
\begin{eqnarray}
\delta\Psi_0(x,\theta)= [\delta x\partial_x+\delta\theta_c\partial^{\ c}_\theta + \delta\theta^c \partial_{\theta c}]\Psi_0(x,\theta), \label{infinitesimalvar}
\end{eqnarray}
which may be put in the form  
\begin{eqnarray}
\delta\Psi_0(x,\theta)= [v\partial_x+\delta\theta_c D^{\ c}_\theta+ \delta\theta^cD_{\theta c}]\Psi_0(x,\theta), \label{supertransf}
\end{eqnarray}
by using the supercovariant derivatives (\ref{covderiv}) and an infinitesimal parameter $v=v(x,\theta)$  given by  
$$
v=\delta x+\theta_c\delta\theta^c+ \theta^c\delta\theta_c.
$$
When the superconformal conditions (\ref{cond2}) are imposed,  the infinitesimally small parameters $\delta x$, $\delta \theta$ and $\delta\theta$ are constrained as
\begin{eqnarray}
D_{\theta a} \delta x&=&\delta \theta_a+\theta_c D_{\theta a}\delta \theta^c =\delta \theta_a- D_{\theta a}(\theta_c\delta \theta^c),  \nonumber\\
D^{\ a}_\theta \delta x&=&\delta \theta^a+\theta^c D^{\ a}_\theta\delta\theta_c=\delta \theta^a- D^{\ a}_\theta(\theta^c\delta \theta_c), \nonumber
\end{eqnarray}
which become respectively 
\begin{eqnarray}
\delta \theta_a={1\over 2}D_{\theta a}v,\quad\quad\quad  
\delta \theta^a={1\over 2}D^{\ a}_\theta v,    \label{Dv}
\end{eqnarray}
by using the chirality condition (\ref{cond1}) in the infinitesimal form 
\begin{eqnarray}
D_{\theta a}\delta\theta_b=D_{\theta a}\delta\varphi_b\Big|_   
{(f,\varphi)=(x,\theta)}=0.    \label{Dtheta}
\end{eqnarray}
 Using this we write  the transformation $\delta \Psi_0$,  given by (\ref{supertransf}),  in a supercovariant form as
\begin{eqnarray}
\delta_v \Psi_0= [v\partial_x+{1\over 2}D_{\theta c}v D^{\ c}_\theta+{1\over 2} D^{\ c}_\theta v D_{\theta c}]\Psi_0  \label{0,0}.  \nonumber
\end{eqnarray}
Hereinafter we do not write the arguments of superfields explicitly if they are simply $(x,\theta)$ as
 here.
This transformation law can be generalized to define a superfield having arbitrary weight $w$  and charge $q$ as
\begin{eqnarray}
\delta_v \Psi_w=  \Big[v\partial_x+{1\over 2}D_{\theta c}v D^{\ c}_\theta+{1\over 2} D^{\ c}_\theta v D_{\theta c}+w\partial_xv+q[D_{\theta c}, D^{\ c}_\theta]v\Big] \Psi_w. \label{h,0}
\end{eqnarray} 
However the charge part of this transformation drops out, since $[D_{\theta c}, D^{\ c}_\theta]v=0$ as discussed in Appendix B.

A posteriori we recognize that 
the superconformal diffeomorphisms (\ref{diffeo}) may be  given  by superfields with weight 0, but  the fermionic ones    
 are constrained by  the chirality condition  
 (\ref{cond1}). That is,   
\begin{eqnarray}
\delta_v f&=&[v\partial_x+{1\over 2}D_{\theta c} v D_{\theta}^{\ c}+ 
{1\over 2}D_{\theta }^{\ c}v D_{\theta c}]f, \label{superconf1} \\
\delta_v \varphi_a&=&[v \partial_x+{1\over 2}D_{\theta c}v D_\theta^{\ c}]\varphi_a,  \label{superconf2}\\
\delta_v \varphi^a&=&[v \partial_x
 +{1\over 2}D_{\theta }^{\ c}v D_{\theta c}]\varphi^a. \label{superconf3}
\end{eqnarray}

We now propose that the $N=4$ super-Schwarzian derivative\footnote{The $N=4$ super-Schwarzian derivative of this form appeared in \cite{Scou}.}
\begin{eqnarray}
{\cal S}(f,\varphi;x,\theta)=\log \det[D_{\theta a}\varphi^b(x,\theta)] \label{Supersch}
\end{eqnarray}
with the above superdiffeomorphisms. When expanded in components by using the formulae in Appendix A, the  purely bosonic part takes the form 
\begin{eqnarray}
{\cal S}(f,\varphi;x,\theta)=\log\partial_xh+{1\over 2}(\theta_a\theta^a)^2\Big[
 -{\partial_x^3h\over\partial_x h}+2({\partial_x^2 h\over\partial_x h})^2
\Big] + O(\eta) , \label{non-susyap}
\end{eqnarray}
in which $h$ is the lowest component of the superfield $f$ and $O(\eta)$ indicates  contributions of 
fermion fields.
The top component   does not coincides with the non-supersymmetric Schwarzian derivative.
 However using this ${\cal S}(f,\varphi;x,\theta)$ we will find  an $N=4$  super-Schwarzian action in the next Section, of which purely bosonic part is the usual non-supersymmetric Schwarzian one. (See (\ref{H}) and the argument thereafter.)  
Or without going through such an argument we may convince ourselves that ${\cal S}(f,\varphi;x,\theta)$ in this form is indeed 
 the $N=4$ super-Schwarzian derivative.  
 Namely  it obeys the anomalous superdiffeomorphism with weight 0 
\begin{eqnarray}
{\cal S}(F(f,\varphi),\Phi(f,\varphi);x,\theta) =
{\cal S}(F,\Phi;f,\varphi)+{\cal S}(f,\varphi;x,\theta), \nonumber
\end{eqnarray}
which  can be easily checked by the chain rule. Infinitesimally it reads
\begin{eqnarray}
\delta_v{\cal S}(f,\varphi;x,\theta)=[v\partial_x+{1\over 2}D_{\theta c} v D_{\theta}^{\ c}+ {1\over 2}D_{\theta }^{\ c}v D_{\theta c}]{\cal S}(f,\varphi;x,\theta)+\partial_x v.  
 \label{superdiff}
\end{eqnarray}
The last term is the conformal anomaly. 

From  (\ref{superdiff}) it follows that   the quantity 
$\det[D_{\theta a}\varphi^b(x,\theta)]$ is a superfield transforming as $\Psi_1$, given by (\ref{h,0}). 
On the other hand we can easily show  that the  quantity 
\begin{eqnarray}
 \Delta\equiv \partial_x f+\varphi_a\partial_x\varphi^a+\varphi^a\partial_x\varphi_a, \label{Delta}
\end{eqnarray}
is also a superfield obeying the same transformation law as $\Psi_1$. 
Hence showing 
\begin{eqnarray}
  \Delta = \det[D_{\theta a}\varphi^b(x,\theta)],
 \label{DDelta}
\end{eqnarray}
 would  give  an alternative check of (\ref{superdiff}). 
The relation (\ref{DDelta}) is indeed proved  in Appendix B as well as 
 $\Delta=\det[D_\theta^{\ a}\varphi_b(x,\theta)]$.

\section{ $N=4$ super-Schwarzian theory}
\setcounter{equation}{0}

Now we are in a position to discuss the coadjoint orbit method to apply for the  $N=4$ 
superconformal algebra. 
The  superconformal algebra $\frak g$  and  the dual space  ${\frak g}^*$ are centrally extended. Their elements  are given by 
\begin{eqnarray}
(u ,k)\in {\frak g}, \quad\quad (b ,c)\in {\frak g}^*.  \nonumber
\end{eqnarray} 
Here $k$ and $c$ are central elements. 
$u$ and $b$ are bosonic superfields,  obeying the superconformal transformations of $\Psi_{-1}$ and $\Psi_0$ given by  (\ref{h,0}) respectively. The transformation of the latter  may be 
centrally extended. 
 The volume element of the $N=4$ superspace, $dxd^4\theta$, has weight 1,
 so that 
   the invariant quadratic form is defined  by 
\begin{eqnarray}
<(b,c),(u,k)>=\int dxd^4\theta\ b u+ck.   \label{invform}
\end{eqnarray}
The  centrally extended superconformal algebra $\frak g$ is given  by  the infinitesimal adjoint action ${\rm ad}(v,l)$ on $(u,k)\in {\frak g}$
\begin{eqnarray}
{\rm ad}(v,l)(u,k) 
&=& \Big( v\partial_x u-u\partial_x v +{1\over 2}D_{\theta c} v D_\theta^{\ c} u +{1\over 2}D^{\ c}_\theta v D_{\theta c} u , \int dxd^4\theta\ v\partial_x u\Big) \nonumber \\
&\equiv& [(u,k),(v,l)].    \label{commut}
\end{eqnarray}
 Then  using  the relation (\ref{invariant}) yields  
  the corresponding coadjoint action ${\rm ad}^*(v,l)$ on $(b ,c)\in {\rm g}^*$  is given by 
\begin{eqnarray}
{\rm ad}^*(v,l)(b,c)=\Big([v\partial_x+{1\over 2}D_{\theta c}v D_\theta^{\ c}+{1\over 2} D_\theta^{\ c} v D_{\theta c}]b+c\partial_x v,0\Big). \label{coadjoint}
\end{eqnarray} 
  We think of a coadjoint orbit O$_{(b,c)}$, whose initial point is $(b,c)\in {\rm g}^*$. 
The finite form of (\ref{coadjoint}) is generated on the coadjoint orbit  by the superdiffeomorphism (\ref{diffeo})
 as\footnote{ We are  sticking to the convention employed below (\ref{Dtheta}), that is, superfields always depend on $(x,\theta,t)$, if the arguments  are not written explicitly.
So this convention is applied to the superfields  $b,f,\varphi_c,\varphi^c$ herein.}  
\begin{eqnarray}
{\rm Ad}^*(f,\varphi)(b,c)&\equiv&\Big(b(f,\varphi)+c{\cal S}(f,\varphi;x,\theta), c\Big).  \label{Ad*}
\end{eqnarray}
Here ${\cal S}(f,\varphi;x,\theta)$ is the super-Schwarzian derivative given by (\ref{Supersch}).

Now the Kirillov-Kostant 2-form (\ref{KK})  can be  given by 
\begin{eqnarray}
\Omega_{(b,c)}={1\over 2}<{\rm Ad}^*(f,\varphi)(b,c),[(y,0),(y,0)]>,  \label{KKK}
\end{eqnarray}
on the coadjoint orbit O$_{(b,c)}$. Here the commutator was given by (\ref{commut}). 
$(y,0)$ is a centrally extended  
 ${\frak g}$-valued 1-form in a space parameterizing the coadjoint orbit. Namely we think of
the superfields $f,\varphi_c,\varphi^c$ in 
 a fictitious space beyond the $D=1, \ N=4$ superspace as  
$f(x,\theta,t_1,t_2,\cdots)$ etc. The 1-form 
$y$ is a function of them. We should have written  it as $y(f,\varphi)$ according to our convention. But we would not like to do it for simplicity hereinafter too.  
$y$ is determined so that  the exterior derivative of the quantity (\ref{Ad*}), which is an element of ${\frak g}^*$, is induced 
by the infinitesimal coadjoint action (\ref{coadjoint}) on it along  the orbit O$_{(b,c)}$  as
\begin{eqnarray}
d{\rm Ad}^*(f,\varphi)(b,c)={\rm ad}^*(y,0)\Big(b(f,\varphi)+c{\cal S}(f,\varphi;x,\theta),c\Big). \label{dY}
\end{eqnarray}
Keep in mind that the exterior derivative acts only on the coordinates $t_1,t_2,\cdots$. 
It is the most important step in our arguments to find an explicit form of $y$ by solving  this equation. It turns out that the solution is given by
\begin{eqnarray}
y={1\over \Delta}(df+ \varphi_c d\varphi^c+\varphi^cd \varphi_c),
 \label{Y}
\end{eqnarray} 
with $\Delta$ defined  by  (\ref{Delta}). 
Once found  $y$ as a solution to (\ref{dY}), the centrally extended  commutator in (\ref{KKK}) 
becomes  
\begin{eqnarray}
[(y,0),(y,0)] =\Big(2y\partial_x y+D_{\theta c} y D_\theta^{\ c} y, \int dxd^4\theta\ y\partial_xy \Big),   \label{yy}
\end{eqnarray}
from  (\ref{commut}).

We shall verify  that  $y$ given by (\ref{Y}) indeed solves the equation (\ref{dY}). 
Using (\ref{coadjoint}) we may rewrite (\ref{dY})  by a pair of the equations 
\begin{eqnarray}
d b(f,\varphi)&=&[y\partial_x+{1\over 2}D_{\theta c}y D_\theta^{\ c}+{1\over 2} D_\theta^{\ c} y D_{\theta c}]b(f,\varphi), \label{db}  \\
d{\cal S}(f,\varphi;x,\theta) &=& [y\partial_x+{1\over 2}D_{\theta c}y D_\theta^{\ c}+{1\over 2} D_\theta^{\ c} y D_{\theta c}]{\cal S}(f,\varphi;x,\theta)+\partial_x y.  \label{dS}
\end{eqnarray}
Compare these with the respective superdiffeomorphisms  
\begin{eqnarray}
\delta_v b(f,\varphi)=[v\partial_x+{1\over 2}D_{\theta c}v D_\theta^{\ c}+{1\over 2} D_\theta^{\ c} v D_{\theta c}]b(f,\varphi) \label{deltab}
\end{eqnarray}
and $\delta_v{\cal S}(f,\varphi;x,\theta)$ given by (\ref{superdiff}). The former one can be verified  by  the infinitesimal variation  (\ref{infinitesimalvar}), in which $\Psi_0(x,\theta)=b(f(x,\theta),\varphi(x,\theta))$. 
The equations (\ref{db}) and (\ref{dS})  require  that  the exterior derivatives of $b(f,\varphi)$ and ${\cal S}(f,\varphi;x,\theta)$ coincide with their  superconfomal diffeomorphisms, if the infinitesimal parameter $v$ is replaced by $y$. In a mathematical language
 we can put it as  
\begin{eqnarray}
i_v d b(f,\varphi)= \delta_v b(f,\varphi), \quad\quad i_vd{\cal S}(f,\varphi; x,\theta)=\delta_v{\cal S}(f,\varphi;x,\theta).  \nonumber
\end{eqnarray}
Here $i_v$ is the anti-derivative of the differential form, implying the operation 
\begin{eqnarray}
i_v d f=\delta_v f, \quad i_v d\varphi_a=\delta_v\varphi_a,\quad i_v d\varphi^a=\delta_v\varphi^a,
 \label{solutions}
\end{eqnarray}
of which the r.h.s.s have been given by (\ref{superconf1})$\sim$(\ref{superconf3}). 
Or  the equation (\ref{dY}) boils down to the following simple equation 
\begin{eqnarray}
i_v  y= v.   \label{boildown}
\end{eqnarray}
The 1-form $y$ given by (\ref{Y}) indeed satisfies this equation by the operation (\ref{solutions}). Thus we have proved that 
it is a right solution for  (\ref{dY}).
The above arguments  might have become too abstract. In Appendix B we show that the equations in (\ref{solutions})
 are obtained from (\ref{Y}) by elementary calculations.

The Kirillov-Kostant 2-form (\ref{KKK}) is invariant under the $N=4$ superdiffeomorphism
by the definition of the quadratic form (\ref{invform}). Therefore we have
$(di_v+i_vd)\Omega_{(b,c)}=0$. $\Omega_{(b,c)}$ is closed so that there exists  a quantity  such as $i_v\Omega_{(b,c)}=dH$.
We shall show that it takes the form 
\begin{eqnarray}
H=\int dxd^4\theta\ v(b(f,\varphi)+c{\cal S}(f,\varphi;x,\theta)),    \label{H}
\end{eqnarray}
with the $N=4$ super-Schwarzian derivative (\ref{Supersch}). To check the claim let us 
 put the Kirillov-Kostant 2-form (\ref{KKK}) in an explicit form  as
\begin{eqnarray}
2\Omega_{(b,c)}=\int dxd^4\theta \Big[(b(f,\varphi)+c{\cal S}(f,\varphi;x,\theta))(2y\partial_x y+D_{\theta c} y D^{\ c}_\theta y)
+cy\partial_x y\Big],  \nonumber
\end{eqnarray}
by (\ref{invform}) with (\ref{Ad*}) and (\ref{yy}). 
Take the anti-derivative and use (\ref{boildown}). By integration by part 
 we get  
\begin{eqnarray}
i_v (2\Omega_{(b,c)})= \int dxd^4\theta 2v d\Big(b(f,\varphi)+c{\cal S}(f,\varphi;x,\theta)\Big),  \nonumber
\end{eqnarray}
owing to 
by (\ref{db}) and (\ref{dS}).  Thus (\ref{H}) has been shown. It is worth knowing about non-supersymmetric approximation
 of ${\cal S}(f,\varphi;x,\theta)$. By using (\ref{non-susyap}) and the expanding formula of  $v$, given  in Appendix A, we find  
the top component of the integrand as  
\begin{eqnarray}
 v{\cal S}(f,\varphi;x,\theta)= \cdots\cdots+{1\over 2}(\theta\cdot\theta)^2\Big\{
 -(\partial^2\alpha)\log \partial_xh+ \alpha\Big[-{\partial_x^3h\over\partial_x h}+2({\partial_x^2 h\over\partial_x h})^2\Big] + O(\eta)\Big\} \nonumber
\end{eqnarray}
Here $\alpha$ is the lowest component of $v$ and $dv=0$. Upon integrating the first term  by part the top component of the integrand becomes the ordinary Schwarzian derivative multiplied by $-2$.  So there is nothing wrong   to have claimed that ${\cal S}(f,\varphi;x,\theta)$ given by (\ref{Supersch}) is  the $N=4$ super-Schwarzian derivative of which purely bosonic part  is given by (\ref{non-susyap}). 
 Finally putting $v=1$  in  (\ref{H})  leads us to the desired super-Schwarzian action.

\section{ PSU(1,1$|$2) symmetry}
\setcounter{equation}{0}

In this section we show symmetry of the acion (\ref{H})  under  PSU(1,1$|$2). 
The action depends on the initial point $b$ of the coadjoint orbit. We discuss the issue dividing the dependence into two cases. For each case we are involved in different realization of the PSU(1,1$|$2) symmetry. 

\vspace{0.5cm}
\noindent
{\bf i)} ${\bf b=0}$.

We expect it to be  realized  on a supermanifold whose local coordinates are the  
 superdiffeomorphism $f,\varphi_a,\varphi^a$ discussed in Section 3 and their complex conjugate  $\overline f,\overline\varphi_a,\overline\varphi^a$. Such a supermanifold is given by 
 the coset space PSU(1,1$|$2)/\{SU(2)$\otimes$U(1)\} for which the generators of PSU(1,1$|$2) are decomposed as
\begin{eqnarray}
\{T^A\}=\{\underbrace{L,F_a,F^a, \overline L, \overline F_a,\overline F^a}_{{{\rm PSU}(1,1|2)\over{\rm SU}(2)\otimes{\rm U}(1)}},\hspace{-0.2cm}\underbrace{L^0,R^a_{\ b}}_{{\rm SU}(2)\otimes{\rm U}(1)}\hspace{-0.2cm}\}.   \label{PSUgene}
\end{eqnarray}
The coset generators $L, F^a,F_a$ correspond to the coordinates  $f,\varphi_a,\varphi^a$. The fermionic coordinates $\varphi_a$ and $\varphi^a$ are doublets of the subgroup SU(2). 
It is well-known that  PSU(1,1$|$2) can be embedded in the larger supergroup D(2,1;$\gamma$).  We may write the fermionic generators and the corresponding coordinates by using the notation of D(2,1;$\gamma$) as
$$
F^{a\dot\alpha}=\left(
\begin{array}{cc}
F^{1\dot 1} &  F^{1\dot 2} \\
F^{2\dot 1} &  F^{2\dot 2} 
\end{array}\right), \quad\quad\quad
\varphi_{a\dot\alpha}=\left(
\begin{array}{cc}
\varphi_{1\dot 1} &  \varphi_{1\dot 2} \\
\varphi_{2\dot 1} &  \varphi_{2\dot 2} 
\end{array}\right).
$$
with the identifications 
\begin{eqnarray}
F^a={1\over \sqrt 2}\left(
\begin{array}{c}
F^{1\dot 1} \\
F^{2\dot 1}
\end{array} 
\right),\quad\quad\quad 
F_a={1\over\sqrt 2}\left(
\begin{array}{c}
\hspace{-0.15cm}\ \ F^{2\dot 2} \\
\hspace{-0.15cm}-F^{1\dot 2}
\end{array} 
\right),    \nonumber
\end{eqnarray}
\vspace{-0.5cm}
\begin{eqnarray}
\varphi_a={1\over \sqrt 2}\left(
\begin{array}{c}
\varphi_{1\dot 1} \\
\varphi_{2\dot 1}
\end{array} 
\right),\ \quad\quad\quad 
\varphi^a={1\over\sqrt 2}\left(
\begin{array}{c}
\hspace{-0.15cm}\ \ \varphi_{2\dot 2} \\
\hspace{-0.15cm}-\varphi_{1\dot 2}    \label{doublet}
\end{array} 
\right).
\end{eqnarray}
See \cite{AH} for the more precise relation between the generators of PSU(1,1$|$2) and D(2,1;$\gamma$). Knowing the Lie algebra of  
D(2,1;$\gamma$) given in a rather simple form,  we can write down that of PSU(1,1$|$2) as  
\begin{eqnarray}
[R^a_{\ b},R^c_{\ d}]&=&-\delta^c_b R^a_{\ d}+\delta^a_d R^c_{\ b}, \hspace{3cm}  [R^a_{\ b},L^0]=0,  \nonumber\\
&\ &\hspace{2.0cm} [\overline L,L]=2L^0,    \nonumber\\
\ [L,L^0]&=&-L,    \hspace{5.5cm}    [\overline L,L^0]=\overline L, \nonumber\\
\ [F^{a\dot\alpha},L^0]&=&-{1\over 2}F^{a\dot\alpha},  \hspace{4.4cm} [\overline F^{a\dot\alpha},L^0]={1\over 2}\overline F^{a\dot\alpha}, \nonumber\\  
\ [F^{a\dot\alpha},L]&=& 0 ,  \hspace{5.7cm} [\overline F^{a\dot\alpha}, L]=F^{a\dot\alpha},     \nonumber\\
\ [F^{a\dot\alpha},\overline L]&=& -\overline F^{a\dot\alpha} ,  \hspace{5.2cm} [\overline F^{a\dot\alpha},\overline L]=0,     \nonumber\\
\ [F^{a\dot\alpha},R^b_{\ c}]&=&\delta^a_c F^{b\dot\alpha}-{1\over 2}\delta^b_c F^{a\dot\alpha}, \hspace{2.7cm} 
 [\overline F^{a\dot\alpha},R^b_{\ c}]=\delta^a_c \overline F^{b\dot\alpha}-{1\over 2}\delta^b_c \overline F^{a\dot\alpha},
 \nonumber\\
\ \{F^{a\dot\alpha},F^{b\dot\beta}\}&=& -\varepsilon^{ab}\varepsilon^{\dot\alpha\dot\beta}L, \hspace{3.7cm} 
\{\overline F^{a\dot\alpha},\overline F^{b\dot\beta}\}= -\varepsilon^{ab}\varepsilon^{\dot\alpha\dot\beta}\overline L,
 \nonumber\\
&\ &\hspace{0cm} \{F^{a\dot\alpha},\overline F^{b\dot\beta}\}=-\varepsilon^{ab}\varepsilon^{\dot\alpha\dot\beta}L^0+
\varepsilon^{ac}\varepsilon^{\dot\alpha\dot\beta}R^b_{\ c}.   \label{LieAlgebra}
\end{eqnarray}
By means of these commutation relations we can calculate the Killing vectors on the coset space PSU(1,1$|$2)/\{SU(2)$\otimes$U(1)\} following the general method developed in \cite{Ao}. They were worked out in \cite{Ho} \footnote{Precisely speaking, it was the Killing vectors of the coset space PSU(2$|$2)/\{SU(2)$\otimes$U(1)\} that were calculated in \cite{Ho}. There use was made of the Lie algebra of PSU(2$|$2), which is given  by (\ref{LieAlgebra}) with $\overline L$ replaced by $-\overline L$. 
 The Killing vectors given below in this paper can be obtained from   
those  given by (2.43) and (2.44) in \cite{Ho} with the replacement $\epsilon_{\overline L}\rightarrow -\epsilon_{\overline L}$.}
\begin{eqnarray}
\delta_\epsilon f &\equiv&-i\epsilon_A R^A       \nonumber\\
  &=& \epsilon_L+f\epsilon_{L^0}+{1\over 2}\varphi_{a\dot\alpha}\epsilon_{Fb\dot\beta}\varepsilon^{ab}\varepsilon^{\dot\alpha\dot\beta}
+{1\over 2}\Big(2f^2\epsilon_{\overline L}+f\varphi_{a\dot\alpha}\epsilon_{\overline Fb\dot\beta}\varepsilon^{ab}\varepsilon^{\dot\alpha\dot\beta}\Big) 
\nonumber\\
&\ & -{1\over 12}\varphi_{a\dot\alpha}\varphi_{b\dot\beta}\varphi_{c\dot\gamma}\epsilon_{\overline Fd\dot\delta}\varepsilon^{cb}\varepsilon^{\dot\gamma\dot\delta}\varepsilon^{ad}\varepsilon^{\dot\alpha\dot\beta}
-{1\over 24}\varphi_{a\dot\alpha}\varphi_{b\dot\beta}\varphi_{c\dot\gamma}\varphi_{d\dot\delta}\epsilon_{\overline L}\varepsilon^{ac}\varepsilon^{\dot\gamma\dot\delta}\varepsilon^{db}\varepsilon^{\dot\alpha\dot\beta}, 
 \label{Killingf}\\
\delta_\epsilon \varphi_{a\dot\alpha} &\equiv&-i\epsilon_A R^A_{\  a\dot\alpha}       \nonumber\\
 &=& \epsilon_{Fa\dot\alpha}+f \epsilon_{\overline Fa\dot\alpha}+{1\over 2}\varphi_{a\dot\alpha}\epsilon_{L^0}
 -\varphi_{b\dot\alpha}\epsilon^b_{R\ a}   \nonumber\\
&\ & +{1\over 2}\Big(2f\varphi_{a\dot\alpha}\epsilon_{\overline L}+\varphi_{b\dot\alpha}\varphi_{c\dot\gamma}
\epsilon_{\overline Fa\dot\beta}\varepsilon^{bc}\varepsilon^{\dot\beta\dot\gamma}\Big)+
{1\over 6}\varphi_{b\dot\alpha}\varphi_{c\dot\gamma}\varphi_{a\dot\beta}
\epsilon_{\overline L}
\varepsilon^{bc}\varepsilon^{\dot\beta\dot\gamma}. \label{Killingphi}
\end{eqnarray}
Here $R^A$ and $R^{A}_{\ a\dot\alpha}$ are the Killing vectors satisfying the Lie algebra of PSU(1,1$|$2). 
 $\epsilon_A$ are infinitesimal parameters of the transformation corresponding to the generators  of PSU(1,1$|$2), given by  (\ref{PSUgene}).

It is not guaranteed at all that the PSU(1,1$|$2) transformations generated by these Killing vectors (\ref{Killingf}) and (\ref{Killingphi}) respect 
the chirality conditions (\ref{cond1}) as well as the superconformal conditions (\ref{cond2}). So we claim that 
\begin{eqnarray}
D_{\theta a}\delta_\epsilon\varphi_b=0, \quad\quad D^{\ a}_\theta\delta_\epsilon\varphi^b=0, \label{chirality1}
\end{eqnarray} 
and
\begin{eqnarray}
D_{\theta a}\delta_\epsilon f&=&\delta_\epsilon\varphi_b D_{\theta a} \varphi^b+\varphi_b D_{\theta a}\delta_\epsilon \varphi^b,
\quad\quad D^{\ a}_\theta\delta_\epsilon f=\delta_\epsilon \varphi^b D^{\ a}_\theta\varphi_b +
\varphi^b D^{\ a}_\theta\delta_\epsilon \varphi_b.     \label{chirality2}
\end{eqnarray}
This claim  will be verified  in Appendix C. Therefore it makes perfect sense to study the transformation property  of the  
 $N=4$ super-Schwarzian action  by the Killing vectors (\ref{Killingf}) and (\ref{Killingphi}). 
Remarkably we find the quantity $\Delta$, given by  (\ref{Delta}),   to obey a fairly simple transformation  as  
\begin{eqnarray}
\delta_\epsilon\Delta = \Big(\epsilon_{L^0}
+2f\epsilon_{\overline L}+\varphi_{c\dot\gamma}\epsilon_{\bar F d\dot\delta}\epsilon^{cd}\epsilon^{\dot\gamma\dot\delta}\Big)\Delta. \label{fff1}
\end{eqnarray} 
This follows by a straightforward calculation with the use of  $\Delta$  written in the notation of $D(2,1;\gamma)$  as
\begin{eqnarray}
\Delta = \partial_xf+{1\over 2}\varphi_{c\dot\gamma}\partial_x\varphi_{ d\dot\delta}\epsilon^{cd}\epsilon^{\dot\gamma\dot\delta}. \nonumber 
\end{eqnarray} 
As the result the  action (\ref{H}) with $b(f,\varphi)=0$ transforms   as
\begin{eqnarray}
\delta_\epsilon H|_{v=1,b=0}=c\int dxd^4\theta \delta_\epsilon\log\Delta
= c\int dxd^4\theta\Big(\epsilon_{L^0}
+2f\epsilon_{\overline L}+\varphi_{c\dot\gamma}\epsilon_{\bar F d\dot\delta}\epsilon^{cd}\epsilon^{\dot\gamma\dot\delta}\Big),
 \label{VariationDelta}
\end{eqnarray}
in which ${\cal S}(f,\varphi;x,\theta)=\log\Delta$
 owing to (\ref{Supersch}) and (\ref{DDelta}). We find that 
the  top component of the integrand  is of the form $\partial_x(\cdots)$, when the superfields  $f$ and $\varphi_{c\dot\gamma}$ are expanded in components as 
in Appendix A and use is made of the second equation in (\ref{constraints}). Therefore the Schwarzian action $H|_{v=1,b=0}$ is invariant 
under the  PSU(1,1$|$2) transformations generated    by the Killing vectors (\ref{Killingf}) and (\ref{Killingphi}).

\vspace{0.5cm}

\noindent
{\bf ii)} ${\bf b\ne0.}$

The infinitesimal parameter $v$ of the $N=4$ superdiffeomorphism  is expanded in components in Appendix A.  The modes  of the  components 
\begin{eqnarray}
\alpha &=& e^{\pm inx}\alpha_{\pm n},\ \  \alpha_0, \label{alpha}\\
 \beta_a &=&  e^{\pm{1\over 2} inx}\beta_{a \pm{1\over 2} n},\label{beta}\\ 
 \beta^a &=&  e^{\pm{1\over 2} inx}\beta^a_{\ \pm{1\over 2} n},\ \   \label{beta'}\\
t^i &=&  t^i_0,    \label{t}
\end{eqnarray}
span the $N=4$ superconformal algebra\cite{Eguchi}.  
The PSU(1,1$|$2) symmetry is realized also by the modes of the diffeomorphisms with $n$ odd. 
They  sequentially correspond  to the generators 
\begin{eqnarray}
&\ &  L,\ \ {\overline L},\ \  L^0 \nonumber\\
&\ & F_a, \ \ {\overline F_a},  \nonumber\\
&\ & F^a, \ \ {\overline F^a},  \nonumber\\
&\ &  R^i,  \nonumber
\end{eqnarray}
 in (\ref{LieAlgebra})\footnote{Note that $R^i= (\sigma^i)^a_{\ b}R^b_{\ a}$.}. 
It is wise to write the Schwarzian action (\ref{H}) as
\begin{eqnarray}
H_{v=1}&=& \int dxd^4\theta\Big(b(f,\varphi)+c{\cal S}(f,\varphi;x,\theta)\Big) \nonumber\\
&\equiv & \int dxd^4\theta\ {\rm Ad}^*(f,\varphi)b(x,\theta).  \label{Ha}
\end{eqnarray}
In the second line we have abused  the definition (\ref{Ad*}) since the $c$-dependence of the initial point of the coadjoint orbit ${\rm O}_{(b,c)}$ is implicit. But  as for the arguments of the initial point $b$ we have made it  explicit  as $b(x,\theta)$ 
 against the convention employed below (3.10). 
 Now the question is if there exists a certain configuration of $b(x,\theta)$ with which the Schwarzian action  is invariant by the superdiffeomorphism given by (\ref{alpha})$\sim$(\ref{t}) with $n$ odd. It may be examined at the initial point of the coadjoint orbit O$_{b,c}$, i.e.,   
\begin{eqnarray}
\Big[\delta_v {\rm Ad}^*(f,\varphi)b(x,\theta)\Big]\Big |_{(f,\varphi)=(x,\theta)}&=&\Big[\delta_v \Big(b(f,\varphi)+c{\cal S}(f,\varphi;x,\theta)\Big)\Big]\Big |_{(f,\varphi)=(x,\theta)} \nonumber\\
&=&v[\partial_x +{1\over 2}D_{\theta a} vD_\theta^{\ a}+{1\over 2}D_\theta^{\ a} vD_{\theta a}]b(x,\theta)+c\partial_x v.
\label{anomalyeq}
\end{eqnarray} 
Here use was made of (\ref{superdiff}) and (\ref{deltab}).

We may proceed the argument quite analogously to   the non-supersymmetric case, but in a much simpler way.  
The Schwarzian action is found 
as
\begin{eqnarray}
H|_{v=1}=\int dx \Big((\partial_x h)^2b(h) +c{\cal S}(h;x)\Big)
\equiv \int dx {\rm Ad}^*(h)b(x).  \label{Hnon-susy} 
\end{eqnarray}
 Having conformal weight 2 the field $b(x)$ gets scaled with a factor $(\partial_x h)^2$
 by the coadjoint action. Assuming $b(x)\ne 0$
we require that 
\begin{eqnarray}
\delta_\alpha{\rm Ad}^*(h)b(x)=[\alpha\partial_x+2\partial_x \alpha]b(x)+c\partial_x^3\alpha
=0 \label{anomalyeq2} 
\end{eqnarray}
under  the diffeomorphism with an infinitesimal parameter $\alpha$. It is important to observe that this is a third-order equation for $\alpha$. If $b(x)$ is constant, then it is solved by any constant $\alpha$. 
 It implies that the action is invariant under U(1) symmetry generated by $L^0$. If
 $b(x)$ is fixed to be  ${1\over 2}cn^2$, then (\ref{anomalyeq2})  admits three independent  solutions of the form (\ref{alpha}).  
The symmetry of the action is enhanced to SL(2). This result is well-known in \cite{Wit2,Bak,Al, Del} as well as \cite{Wi}. For the case of $b(x)=0$ refer to a comment in the end of the paper. 

 Let us turn to the $N=4$ Schwarzian action (\ref{Ha}). 
The superfield $b(x,\theta)$
is expanded  in components as the superfield $f$ was done in Appendix A, i.e.,  
\begin{eqnarray}
b(x,\theta)&=& a+\theta\cdot \gamma+\gamma\cdot\theta + \theta\cdot\theta i
+(\theta\sigma^i\theta)s^i \nonumber\\
&+&{1\over 2}\epsilon_{ab}\theta^a\theta^bj+{1\over 2}\epsilon^{ab} \theta_a\theta_bk + (\theta\cdot\theta)(\theta\cdot\sigma)+(\theta\cdot\theta)(\sigma\cdot\theta)
+(\theta\cdot\theta)^2d.
\label{expansionb}
\end{eqnarray}
Here the arguments of the component fields have  been omitted according to our convention. 
 Put this expansion as well as that of $v$, also given in Appendix A,  into  the second line of (\ref{anomalyeq}).  Calculating its top component we have\footnote{Our convention is that $\int d^4\theta (\theta\cdot\theta)^2 =2$} 
\begin{eqnarray}
\Big[\delta_vH_{v=1}\Big]\Big|_{(f,\varphi)=(x,\theta)}&=& \int dx\Big\{2\partial_x d\alpha+4d\partial_x\alpha+\partial_xa\partial_x^2\alpha
-c\partial_x^3\alpha   \nonumber\\
&\ &\hspace{-0.8cm} +{1\over 2}\Big(\hspace{-2mm}-(\partial_x\sigma\cdot\beta)-(\partial_x\gamma\cdot\partial_x\beta)
-3(\sigma\cdot\partial_x\beta)+(\gamma\cdot\partial_x^2\beta)
\nonumber\\
&\ &\hspace{0cm} +(\beta\cdot\partial_x\sigma)
-(\partial_x\beta\cdot\partial_x\gamma)+3(\partial_x\beta\cdot\sigma)
+(\partial_x^2\beta\cdot\gamma)\Big) -4s^i\partial_x t^i\Big\}.
\nonumber
\end{eqnarray}
We find that it is vanishing by the diffeomorphism (\ref{alpha})$\sim$(\ref{t}) when 
the initial point $b$ has a configuration such as
\begin{eqnarray}
&\ & a=0, \quad d=-{1\over 4}cn^2 \quad s^i=s^{\hspace{2mm}i}_0,\nonumber\\
&\ &  \gamma_a=e^{\pm{\sqrt 3\over 2} nx}\gamma_{a0},\quad \gamma^a=e^{\pm{\sqrt 3\over 2} nx}\gamma^{a}_{\ 0},
\label{solution}
\end{eqnarray}
with
\begin{eqnarray}
\sigma_a=-{1\over 3}\partial_x\gamma_a ,\quad \sigma^a={1\over 3}\partial_x\gamma^a. 
\nonumber
\end{eqnarray}
Thus the Schwarzian action is invariant under PS(1,1$|$2). 
But it is worth remarking that we do not encounter 
 boundary terms  at all in examining the symmetry of the integrand. 
It is also worth recognize that the solution contains the non-supersymmetric one in the previous paragraph by setting $2d=-b$. 

The reader may  ask about symmetry for the density of the Schwarzian action (\ref{Ha}). Then the variation  (\ref{anomalyeq}) is required to vanish at lower orders of $\theta$ as well. 
The resulting differential equations are too stringent  to be satisfied by the above solution.  For instance at the lowest order of 
 $\theta$ it reads
\begin{eqnarray}
\partial_x a\alpha+c\partial_x\alpha +{1\over 2}(\beta\cdot\gamma)+{1\over 2}(\gamma\cdot\beta)=0. 
\nonumber
\end{eqnarray} 
More stringent equations come out at higher orders. Nonetheless it is not hard to
 see that all the equations are satisfied  by the subset of the modes 
$$ 
\alpha=\alpha_0, \quad\quad \beta_a=\beta^a=0, \quad\quad t^i=t^{i}_{\hspace{1mm}0}, 
$$
in  (\ref{alpha})$\sim$(\ref{t}),  
when  $b$  has a configuration such as 
\begin{eqnarray}
 d=d_0, \quad\quad others = 0.
\nonumber
\end{eqnarray}
Therefore  the subgroup
SU(2)$\otimes$U(1) is also a symmetry of the density of the Schwarzian action (\ref{Ha}).

\vspace{0.5cm}

 We content ourselves with these solutions, although
our analysis of the differential equations is not exhaustive at all. In summary, the partition function of the $N=4$ super-Schwarzian theory is given by 
$$
Z=\int_{\cal M}{\cal D}f{\cal D}\varphi_a{\cal D}\varphi^a \exp\Big(H|_{v=1}\Big),
\quad\quad {\cal M}={\rm superdiff}/{\rm PSU}(1,1|2),  
$$
when the action is symmetric under PSU(1,1$|$2). 

\section{Conclusions}

In this paper we have formulated an $N=4$ super-Schwarzian action by means of the coadjoint
orbit method. The  action is dependent on the initial point $b$ of the orbit. For the case of 
$b=0$ it has been shown to have  symmetry under PSU(1,1$|$2) realized by the Killing vectors for the coset space PSU(1,1$|$2)/\{SU(2)$\otimes$U(1)\}.  
When  $b\ne 0$ we have also shown that it becomes invariant by  a  set of modes of the superdiffeomorphism realizing   PSU(1,1$|$2). 
 For that we have found a configuration of $b$ such as  given by (\ref{solution}).

 We comment  the case of  $b=0$ for the non-supersymmetric Schwarzian action (\ref{Hnon-susy}), which we have not discussed in Section 5. The non-supersymmetric Schwarzian derivative   
 ${\cal S}(h;x)$  is  invariant  under  SL(2)  realized by the Killing vectors for the coset space SL(2)/U(1).  
However  (\ref{VariationDelta}) implies that the  Schwarzian derivative ${\cal S}(f,\varphi;x,\theta)$ for the  $N=4$ case is invariant  only modulo  boundary terms $\partial_x(\cdots)$ by the same transformation.
This discrepancy is not a problem because  the purely bosonic part of ${\cal S}(f,\varphi;x,\theta)$ is given by (\ref{non-susyap}) and the top component giving the action
 reads 
$$
-{\cal S}(h;x)+{1\over 2}({\partial^2_x h\over \partial_x h})^2.
$$   
The additional term is invariant modulo the boundary term $2\epsilon_{\bar L}\partial^2_x h$ under SL(2)  
 realized by the Killing vectors.  It is consistent with (\ref{VariationDelta}).

It is desirable to study quantum dynamics of the $N=4$ super-Schwarzian action. Our study on this is in progress. 
 It is also desirable  to extend  the $D=2$ Liouville gravity  to the $N=4$ supersymmetric  
 one. It will be reported in \cite{A5}.

\vspace{0.5cm}

\appendix

\section{Superfields in components } 
\setcounter{equation}{0}

In the body of the paper the $N=4$ super-Schwarzian derivative $\cal S$ was needed to be expanded in components.
 We give here only the expansion for the basic ones.
The superfields $f, \varphi_c, \varphi^c$ which describe the $N=4$ superdiffeomorphism  are expanded as
\begin{eqnarray}
 f(x,\theta)&=& h(x)+\theta\cdot\psi(x)+\psi(x)\cdot\theta+\theta\cdot\theta l(x)+(\theta\sigma^i\theta)t^i(x)    \nonumber\\
 &+&{1\over 2}\epsilon_{ab}\theta^a\theta^b m(x) + {1\over 2}\epsilon^{ab}\theta_a\theta_b n(x)+
(\theta\cdot\theta)(\theta\cdot\omega(x))+(\theta\cdot\theta)(\omega(x)\cdot\theta)+(\theta\cdot\theta)^2g(x),        
 \nonumber\\
\varphi_c(x,\theta) &=& \rho(x+\theta\cdot\theta)\Big[\theta_c+\eta_c(x+\theta\cdot\theta)+{1\over 2}\epsilon^{ab}
\theta_a\theta_b\alpha_c(x)(x+\theta\cdot\theta)\Big], \nonumber\\
\varphi^c(x,\theta)&=& \xi(x-\theta\cdot\theta)\Big[\theta^c+\eta^c(x-\theta\cdot\theta)+{1\over 2}\epsilon_{ab}\theta^a\theta^b\alpha^c(x-\theta\cdot\theta)\Big],     \nonumber
\end{eqnarray}
with $\theta\cdot\psi\equiv\theta_a\psi^a$, $\psi\cdot\theta\equiv\psi_a\theta^a$  etc. Note that the component fields  of $\varphi_c$ and $\varphi^c$  got 
 the argument $x$ shifted so that the chirality conditions (\ref{cond1}) are satisfied. By imposing
the superconformal conditions (\ref{cond2}) they become
\begin{eqnarray}
 f(x,\theta)&=&h+\rho\xi\Big[\theta\cdot\eta-\eta\cdot\theta\Big] \nonumber\\
  &+&\theta\cdot\theta
\partial_x\Big(\rho\eta\cdot\xi\eta\Big)+2{\xi\over \rho}\Big(\rho\eta\cdot\theta\Big)\Big(\partial_x(\rho\eta)\cdot\theta\Big)
+2{\rho\over \xi}\Big(\theta\cdot\xi\eta\Big)\Big(\theta\cdot\partial_x(\xi\eta)\Big)  \nonumber\\
  &+&\theta\cdot\theta\Big[\Big(\theta\cdot\partial_x(\rho\xi\eta)\Big) +\Big(\partial_x(\xi\rho\eta)\cdot\theta\Big)\Big]    \nonumber\\
&+&{1\over 2}(\theta\cdot\theta)^2\Big[-\Big(\rho\eta\cdot\partial_x^2(\xi\eta)\Big)+\Big(\partial_x^2(\rho\eta)\cdot\xi\eta\Big)
 +\xi\partial_x\rho+\rho\partial_x\xi\Big], \nonumber\\
\varphi_c(x,\theta) &=& \rho\eta_c+\rho\theta_c+\theta\cdot\theta\partial_x(\rho\eta_c)
+ {1\over 2}\epsilon^{ab}\theta_a\theta_b\rho\alpha_c+
\theta\cdot\theta\theta_c\partial_x\rho +{1\over 2}(\theta\cdot\theta)^2\partial_x^2(\rho\eta_c), \nonumber\\
\varphi^c(x,\theta)&=&\xi\eta^c+\xi\theta^c-\theta\cdot\theta\partial_x(\xi\eta^c)+
{1\over 2}\epsilon_{ab}\theta^a\theta^b\xi\alpha^c-\theta\cdot\theta\theta^c\partial_x\xi
+{1\over 2}(\theta\cdot\theta)^2\partial_x^2(\xi\eta^c),\nonumber
\end{eqnarray}
with the remaining constraints 
\begin{eqnarray}
&\ &\hspace{-2cm}\quad \xi\partial_x\rho=\rho\partial_x\xi,\quad 
 \partial_x h+\Big(\rho\eta\cdot\partial_x(\xi\eta)\Big)-\Big(\partial_x(\rho\eta)\cdot\xi\eta\Big)=\rho\xi,
\nonumber\\
&\ &\hspace{-2cm}\quad \xi\alpha_a=2\epsilon_{ab}\partial_x(\xi\eta^b),\quad\quad \rho\alpha^a=2\epsilon^{ab}\partial_x(\rho\eta_b).  \label{constraints}
\end{eqnarray}
Now the component fields  have the argument $x$, which has been omitted for simplicity.
 It is important to note that 
 all of their  top components  are  of the form $\partial_x
(\cdots)$. Use the second equation of (\ref{constraints}) in order to see this for the one of $f$.

In the end of Section 4 we also need to expand the infinitesimal parameter $v(x,\theta)$ of the $N=4$ superdiffeomorphism. 
By (\ref{Dv}) and (\ref{Dtheta}) it satisfies 
$$
D_{\theta a}D_{\theta b}v=0, \quad\quad D_\theta^{\ a}D_\theta^{\ b}v=0,
$$
so that 
\begin{eqnarray}
v(x,\theta)=\alpha+\theta\cdot\beta- \beta\cdot\theta+(\theta\sigma^i\theta)t^i-
(\theta\cdot\theta)(\theta\cdot\partial_x\beta)-(\theta\cdot\theta)(\partial_x\beta\cdot\theta)
-{1\over 2}(\theta\cdot\theta)^2\partial_x^2\alpha, \nonumber
\end{eqnarray}
in which $\alpha$, $\beta_a$, $\beta^a$, $t^i$ are independent parameters of the superdiffeomorphisms.

\section{Proofs of  some formulae in Sections 3 and 4} 
\setcounter{equation}{0}

We prove the various formulae required for the arguments in Sections 3 and 4. 
 We begin by the following formulae
\begin{eqnarray}
(D_{\theta a}\varphi^c)(D_\theta^{\ b}\varphi_c)&=&\delta^b_a  
(\partial_x f+ \varphi_c \partial_x\varphi^c+\varphi^c \partial_x\varphi_c)\equiv\delta^b_a\Delta, \label{f1} \\
(D_\theta^{\ c}\varphi_a)(D_{\theta c}\varphi^b)&=& \delta^b_a 
(\partial_x f+ \varphi_c \partial_x\varphi^c+\varphi^c \partial_x\varphi_c)\equiv\delta^b_a\Delta, \label{f1'} \\
 (D_{\theta a}\varphi^b)(D^{\ a}_\theta\varphi_b)&=&2(\partial_x f+ \varphi_c \partial_x\varphi^c+\varphi^c \partial_x\varphi_c) \equiv 2\Delta,  \label{f2}\\ 
 2D_{\theta a}\varphi^b\partial_x \varphi_b&=&D_{\theta a}\Delta, \quad\quad
2D^{\ a}_\theta\varphi_b\partial_x \varphi^b=D^{\ a}_\theta\Delta,  \label{f4} \\
 \det[D_{\theta a}\varphi^b]\det[D^{\ a}_\theta\varphi_b]&=& \Delta^2. \label{f3}
\end{eqnarray}
They were studied in \cite{Ue}. 
(\ref{f1'}) and (\ref{f2}) follow from (\ref{f1}).  (\ref{f1}) can be shown by 
taking the supercovariant derivative of (\ref{cond2}) and using the algebra (\ref{algebra}) and the chirality condition (\ref{cond1}) as 
\begin{eqnarray}
D_\theta^{\ b}D_{\theta a}f&=&(D_\theta^{\ b}\varphi_c)(D_{\theta a}\varphi^c)-2\delta^b_a\varphi_c\partial_x\varphi^c, \nonumber\\
D_{\theta a}D_\theta^{\ b}f&=&(D_{\theta a}\varphi^c)(D_\theta^{\ b}\varphi_c)-2\delta^b_a\varphi^c\partial_x\varphi_c. \nonumber
\end{eqnarray}
(\ref{f4}) can be shown by similarly taking the supercovariant derivative of (\ref{f2}). 
Then calculate the terms $(D_{\theta c}D_{\theta a}\varphi^b)(D^{\ a}_\theta\varphi_b) $ or
$(D_{\theta a}\varphi^b)(D_{\theta}^{\ c}D^{\ a}_\theta\varphi_b) $ in the resulting equation as 
\begin{eqnarray}
(D_{\theta c}D_{\theta a}\varphi^b)(D^{\ a}_\theta\varphi_b) 
&=&-D_{\theta c}\Delta+ 4(D_{\theta c}\varphi^b)\partial_x\varphi_b, \nonumber \\
(D_{\theta a}\varphi^b)(D_{\theta}^{\ c}D_{\theta}^{\ a}\varphi_b) 
&=&-D_{\theta}^{\ c}\Delta+ 4(D_{\theta}^{\ c}\varphi^b)\partial_x\varphi_b, \nonumber
\end{eqnarray}
by the successive use of (\ref{algebra}), (\ref{f1}) and  (\ref{cond1}). We then get (\ref{f4}).
  (\ref{f3}) is now obvious from (\ref{f1}) and (\ref{f1'}). It can be factorized 
to become 
\begin{eqnarray}
\Delta=\det[D_{\theta a}\varphi^b]=\det[D^{\ a}_\theta\varphi_b]. \label{**}
\end{eqnarray}
We have  checked this identity in components by using the expansion formulae  in Appendix A. 

A direct calculation  shows that  the quantity $\det[D_{\theta a}\varphi^b]$ transforms 
by the superconformal transformations (\ref{superconf1})$\sim$(\ref{superconf3}) as 
\begin{eqnarray}
\delta_v \log\det[ D_{\theta a}\varphi^b]
&=&[v\partial_x+{1\over 2}D_{\theta c}\xi D_\theta^{\ c}+
 {1\over 2}D_{\theta}^{\ c}v D_{\theta c}]\log\det[ D_{\theta a}\varphi^b] \nonumber\\
&\ &  +{1\over 2}D_{\theta c} D^{\ c}_\theta v.
  \nonumber
\end{eqnarray}
Similarly we can show that the quantity $\Delta$, defined by (\ref{f1}), transforms as a superfield $\Psi_1$ given by 
 (\ref{h,0}). Both quantities should transform in the same way. Therefore the relation (\ref{**}) implies that  $[D_{\theta c}, D^{\ c}_\theta]v=0$.

By using above  formulae we can prove (\ref{solutions}). Suppose that $y$ is given by (\ref{Y}) and
 take the supercovariant derivative of it.  We then get
\begin{eqnarray}
\Delta D_{\theta a}y=-yD_{\theta a}\Delta +D_{\theta a}(df+ \varphi_c d\varphi^c+\varphi^cd \varphi_c). \label{Dy}
\end{eqnarray}
Calculate the second term in the r.h.s. as 
\begin{eqnarray}
D_{\theta a}(df+ \varphi_c d\varphi^c+\varphi^cd \varphi_c)=2D_{\theta a}\varphi^cd\varphi_c,
 \nonumber
\end{eqnarray}
by  (\ref{cond1}) and (\ref{cond2}). Put this into (\ref{Dy}) and contract  both sides with $D_\theta^{\ a}\varphi_b$.  
 Using (\ref{f1}) and (\ref{f4}) we then find 
$$
d \varphi_a=[y \partial_x+{1\over 2}D_{\theta c}y D_\theta^{\ c}]\varphi_a.
$$
For $d\varphi^a$ the analogous formula can be shown. Substitute $d\varphi_a$ and $d\varphi^a$ in (\ref{Y})   
 by these formulae. 
  We  solve the resulting equation for $df$
  using the superconformal conditions (\ref{cond2}).  
The solution is 
$$
d f=[y\partial_x+{1\over 2}D_{\theta c}y D_\theta^{\ c}+
 {1\over 2}D_\theta^{\ c} y D_{\theta c}]f.
$$
Thus all of the equations in (\ref{solutions}) have been proved.

\section{Proof of (\ref{chirality1}) and (\ref{chirality2}) in Section 5 } 
\setcounter{equation}{0}

We show the formulae  (\ref{chirality1}) and  (\ref{chirality2}) following from the chirality and superconformal conditions respectively.
To this end it is convenient to write the Killing vectors in the doublet notation
\begin{eqnarray}
\delta_\epsilon f &\equiv&-i\epsilon_A R^A     \nonumber \\
  &=& \epsilon_L+f\epsilon_{L^0}+(\varphi_c\epsilon_F^{\  c}+\varphi^c\epsilon_{F c})
+\Big(f^2\epsilon_{\overline L}+f(\varphi_c\epsilon_{\overline F}^{\  c}+\varphi^c\epsilon_{\overline  F c})\Big)  \nonumber\\
&\ & +(\varphi_b\varphi^b)(\varphi_c\epsilon_{\overline F}^{\ c}-\varphi^c\epsilon_{\overline Fc})
+(\varphi_c\varphi^c)^2\epsilon_{\overline L},  \label{Killing2a}  \\
\delta_\epsilon \varphi_{a} &\equiv&-i\epsilon_A R^A_{\  a} \nonumber \\
 &=& \epsilon_{Fa}+f \epsilon_{\overline Fa}+{1\over 2}\varphi_a\epsilon_{L^0}-\varphi_c \epsilon_{R\ a}^{\ c}\nonumber \\ 
&\ & +\Big(f\varphi_a\epsilon_{\overline L}+(\varphi_c\varphi^c\epsilon_{\overline Fa}
+2\varphi_c\epsilon_{\overline F}^{\  c}\varphi_a)\Big)
+\varphi_c\varphi^c\varphi_a\varepsilon_{\overline L},       \label{Killing2b} \\
\delta_\epsilon \varphi^{a} &\equiv&-i\epsilon_A R^{A a}  \nonumber \\
 &=& \epsilon_{F}^{\ a}+f \epsilon_{\overline F}^{\ a}+{1\over 2}\varphi^a\epsilon_{L^0}+\varphi^c\epsilon_{R\ c}^{\ a}\nonumber \\ 
&\ & +\Big(f\varphi^a\epsilon_{\overline L}- (\varphi_c\varphi^c\epsilon_{\overline F}^{\ a}
-2\varphi^c\epsilon_{\overline Fc}\varphi^a)\Big)
-\varphi_c\varphi^c\varphi^a\varepsilon_{\overline L}.   \label{Killing2c}
\end{eqnarray} 
Here we have used the same doublet notation also for $\epsilon_{Fa\dot\alpha}$, $\epsilon_{\overline Fa\dot\alpha}$ and $R^A_{\ a\dot\alpha}$ as given for $\varphi_{a\dot\alpha}$ by (\ref{doublet}). Then it is immediate  to see that (\ref{chirality1}) holds owing to (\ref{cond1}) and  (\ref{cond2}).  (\ref{chirality2}) can be  also checked by a few of calculations. 
 We do it explicitly for  the first equation of  (\ref{chirality2})  as an example. From (\ref{Killing2c})
 it follows that 
\begin{eqnarray}
D_{\theta a}\delta_\epsilon\varphi^b  
&\ &=2D_{\theta a}f \epsilon_{\overline F}^{\ b}+{1\over 2}D_{\theta a}\varphi^b\epsilon_{L^0}
+D_{\theta a}\varphi^c\epsilon_{R\ c}^{\ b} \nonumber\\
&\ &+\Big(2D_{\theta a}f\varphi^b+fD_{\theta a}\varphi^b-\varphi_c\varphi^cD_{\theta a}\varphi^b\Big)\varepsilon_{\overline L}
+2D_{\theta a}(\varphi^c\epsilon_{\overline Fc}\varphi^b),  \label{Ddelta}
\end{eqnarray}
by using (\ref{cond1}) and (\ref{cond2}). 
For the case of $\epsilon=\epsilon_{\overline  F}$  we have 
$$
\varphi_bD_{\theta a}\delta_\epsilon\varphi^b\Big|_{\epsilon=\epsilon_{\overline  F}} =-2D_{\theta a}f\cdot\varphi_b \epsilon_{\overline F}^{\ b}
-2D_{\theta a}\varphi^c\epsilon_{\overline F  c}\cdot\varphi_b\varphi^b
+2\varphi^c\epsilon_{\overline F c}\cdot\varphi_b D_{\theta a}\varphi^b.
$$
By the same calculation  we have also
\begin{eqnarray}
\delta_\epsilon\varphi_b D_{\theta a}\varphi^b\Big|_{\epsilon=\epsilon_{\overline  F}} &=& \Big((f+\varphi_c\varphi^c)\epsilon_{\overline Fb} 
+2\varphi_c\epsilon_{\overline F}^{\ c}\cdot\varphi_b\Big)D_{\theta a} \varphi^b,      \nonumber     \\  
D_{\theta a}\delta_\epsilon f\Big|_{\epsilon=\epsilon_{\overline  F}} &=&2\varphi_bD_{\theta a}\varphi^b\cdot\varphi^c\epsilon_{\overline F c}
+fD_{\theta a}\varphi^c\epsilon_{\overline F c}-\varphi_b\varphi^b\cdot D_{\theta a}\varphi^c\epsilon_{\overline F c},     \label{deltaphiD}
\end{eqnarray}
from (\ref{Killing2b}) and (\ref{Killing2a}) respectively.   
It is now clear  that  the first relation of  (\ref{chirality2})  is satisfied for the case of  $\epsilon=\epsilon_{\overline  F}$.  It can be checked similarly for other cases than $\epsilon=\epsilon_{\overline  F}$.

\end{document}